\documentclass[aps,pra,showpacs,groupedaddress]{revtex4}
\usepackage{amsfonts}
\usepackage{amsmath}
\usepackage{mathrsfs}
\usepackage{amssymb}
\usepackage[dvips]{graphics,color}
\usepackage{graphicx}
\usepackage{dcolumn}
\usepackage{bm}

\begin{document}

\title{Quantum Fluctuations of Vortex-Lattice State in Ultrafast Rotating
Bose Gas}

\author{Qiong Li }

\affiliation{Department of Physics, Peking University, Beijing,
100871, China}

\author{Bo Feng }

\affiliation{Department of Physics, Peking University, Beijing,
100871, China}

\author{Dingping Li}

\affiliation{Department of Physics, Peking University, Beijing,
100871, China}
\begin{abstract}
Quantum fluctuations in an ultrafast rotating Bose gas at zero
temperature are investigated. We calculate the condensate density
perturbatively to show that no condensate is present in the
thermodynamic limit. The excitation from Gaussian fluctuations
around the mean field solution causes infrared divergences in loop
diagrams, nevertheless, in calculating the atom number density, the
correlation functions and the free energy, we find the sum of the
divergences in the same loop order vanishes and obtain finite
physical quantities. The long-range correlation is explored and the
algebraic decay exponent for the single-particle correlation
function is obtained. The atom number density distribution is
obtained at the one-loop level, which illustrates the quantum
fluctuation effects to melt the mean field vortex-lattice. By the
non-perturbative Gaussian variational method, we locate the spinodal
point of the vortex-lattice state.
\end{abstract}

\pacs{03.75.Hh, 03.75.Lm, 05.30.Jp, 05.30.Rt}

\maketitle

\section{introduction}

The appearance of vortex excitations in response to rotation is a
characteristic feature of superfluid \cite{key-1,key-2,key-3}. Since
the discovery of Bose-Einstein condensation (BEC) in atomic gases
\cite{key-4,key-5,key-6}, much work has been devoted to the
properties of rotating gaseous condensates in traps, and these
developments have been reviewed in \cite{key-7,key-8}. As the
rotation frequency $\Omega$ increases, more and more vortices occur,
from a single one to several ones, then they form an Abrikosov
lattice, i.e., a triangular array with a surface density
$n_{\upsilon}=\frac{m\Omega}{\pi\hbar}$($m$ is the mass of the
condensed atoms)
\cite{key-9,key-10,key-11,key-12,key-13,key-14,key-15,key-16,key-17,key-18}.
In the frame co-rotating with harmonic traps, rotation effects are
described by a centrifugal force, which effectively reduces the
transverse harmonic trapping, and a Coriolis force which has the
same mathematical structure as the Lorentz force that an electron
experiences in a uniform magnetic field. In the fast rotation
regime, when the centrifugal force almost cancels the transverse
confinement, the energy levels of the single particle part of the
Hamiltonian organize into Landau levels with spacing $2\hbar\Omega$,
and the condensates expand radically, leading to a very dilute atom
number density, which ensures the mean interaction energy smaller
than Landau level spacing, so that the cold atoms are confined in
the lowest Landau level (LLL) single particle orbit.

For rotating bosonic atoms in the LLL, the filling fraction, i.e.,
the ratio of the number of atoms to the number of vortices, is the
parameter controlling the nature of the system. At high filling
fractions, the condensate is in the mean field quantum-Hall regime
and forms an ordered vortex lattice ground state
\cite{key-19,key-20,key-21,key-22,key-23,key-24}. As the filling
fraction decreases, there is a zero temperature phase transition
from a triangular vortex-lattice to strongly correlated
vortex-liquid\cite{key-25} and the melting point is located by
various approaches \cite{key-25,key-26,key-27} to be approximately
at the filling fraction $6\sim10$. The experiment done by V.
Schweikhard etc. \cite{key-28} created an ordered vortex lattice in
the mean-field quantum-Hall regime, and provided evidence that the
elastic shear strength of the vortex lattice drops substantially as
the BEC enters the mean field quantum-Hall regime.

In the ultrafast rotation limit when the transverse confinement is
exactly canceled by the centrifugal force, the condensates expand to
be a two dimensional configuration and atoms are frozen in the
lowest energy level in the $z$ direction (assuming a strong
confinement in the $z$ direction). For such a two dimensional
system, J. Sinova etc. \cite{key-26} find that the solution to the
Gross-Pitaevskii (GP) equation is an Abrikosov triangular
vortex-lattice and the integral for the fraction of atoms outside
the condensate diverges logarithmically with the system size, which
implies that no BEC occurs in the thermodynamic limit even at zero
temperature.

Fluctuation effects in the transverse plane in an ultrafast rotation
limit are important and one has to go beyond mean field treatment
\cite{key-19}. In the present paper, we are devoted to quantum
fluctuations in the ultrafast rotation limit and focus on the
thermodynamic limit at zero temperature.

Based on analytical calculations in the perturbative framework, we
show that the condensate density is zero in the thermodynamic limit,
namely, there is no BEC. The single-particle correlation function
and the density fluctuation correlation function are shown to fall
off as an inverse power of the separation distance in the large
distance limit, which indicates an algebraic long-range order and
the algebraic decay exponent is obtained. The atom number density
distribution is obtained at the one-loop level, which illustrates
the quantum fluctuation effects to melt the mean field
vortex-lattice.

By loop expansion around the mean field solution, we calculate the
free energy density up to two-loop. The mean field solution is an
Abrikosov triangular vortex-lattice and the excitation from Gaussian
fluctuations has a quadratic dispersion at small wave vectors
\cite{key-26}. We find that the quadratic dispersion causes infrared
divergences in the two-loop diagrams, nevertheless, the sum of the
divergences vanishes and the two-loop contribution to the free
energy density is finite.

We also study the model by non-perturbative Gaussian variational
method. The free energy density calculated by the perturbation
theory and that by the Gaussian variational method coincide very
well at large filling fractions. The vortex lattice solution exists
only when the filling fraction $\nu$ is greater than a certain value
$\nu_{s}$, which is found to be about $1.1$. The point $\nu=\nu_{s}$
is the so called spinodal point. Between the spinodal point and the
melting point is the meta-stable vortex-lattice state.

The paper is organized as follows:

In section \ref{sec:The-Model}, we formulate the model. In section
\ref{sec:long-range correlation}, we explore the long-range
correlations. In section \ref{perturbation theory}, we calculate the
free energy density perturbatively up to two-loop. In section
\ref{sec:Gaussian variational calculation}, we use the
non-perturbative Gaussian variational method to study the model. In
section \ref{sec:Summary}, we give a summary and the conclusions.

\section{The Model\label{sec:The-Model}}

For a system of $N$ bosonic atoms in an axisymmetric harmonic trap
(with trap frequencies $\omega_{\perp}$and $\omega_{z}$) rotating
with angular velocity $\Omega\mathbf{e}_{z}$, the Hamiltonian in the
rotating frame is

\begin{equation}
H=\sum_{i=1}^{N}[\frac{(\mathbf{p}_{i}-m\Omega\hat{z}\times\mathbf{r}_{i})^{2}}{2m}+\frac{1}{2}m(\omega_{\perp}^{2}-\Omega^{2})(x_{i}^{2}+y_{i}^{2})+\frac{1}{2}m\omega_{z}^{2}z_{i}^{2}]+g\sum_{i<j=1}^{N}\delta(\mathbf{r}_{i}-\mathbf{r}_{j}),\end{equation}
where $g=\frac{4\pi\hbar^{2}a_{s}}{m}$ is the strength of the hard
core repulsive interactions, with $a_{s}$ the s-wave scattering
length. The centrifugal force effectively reduces the radial
confinement and the Coriolis force is equivalent to the Lorentz
force exerted on a particle with charge $Q$ by a magnetic field
$\mathbf{B}=\frac{2m\Omega}{Q}\mathbf{e}_{z}$. In the present paper
we are confined to the ultrafast rotation limit by setting
$\Omega=\omega_{\perp}$, and assume the axial confinement is so
strong that atoms are frozen in the lowest harmonic state in the $z$
direction. Consequently, what we concern is essentially a two
dimensional system of charged bosonic atoms experiencing an
effective magnetic field in the $z$ direction, described by the
Hamiltonian
\begin{equation}
H=\sum_{i=1}^{N}\frac{(\mathbf{p}_{i}-m\Omega\hat{z}\times\mathbf{r}_{i})^{2}}{2m}+g\sum_{i<j=1}^{N}\delta(\mathbf{r}_{i}-\mathbf{r}_{j}).\label{eq:H}\end{equation}
The kinetic part of the Hamiltonian has equally spaced Landau levels
(with spacing $2\hbar\Omega$) and the interaction part is a small
perturbation, $ng\ll2\hbar\Omega$ ($n$ is the mean number density of
the atoms). Without thermal fluctuations at zero temperature, all
atoms are confined in the LLL. In the LLL subspace, the kinetic part
of the Hamiltonian is quenched and represented by $\hbar\Omega$,
hence we have the grand Hamiltonian in second quantized form (with
the unit $\hbar=1$)\begin{equation} \hat{H}-\mu\hat{N}=\int
d^{2}\mathbf{r}\left[(\Omega-\mu)\Psi^{\dagger}(\mathbf{r})\Psi(\mathbf{r})+\frac{g}{2}\Psi^{\dagger}(\mathbf{r})\Psi^{\dagger}(\mathbf{r})\Psi(\mathbf{r})\Psi(\mathbf{r})\right],\end{equation}
and the grand-canonical partition function in the functional
formalism
\begin{equation}
\mathcal{Z}(\beta,\mu)=\int\mathcal{D}[\Psi^{\ast}\Psi]e^{-S[\Psi^{\ast},\Psi]},\end{equation}
where the action $S[\Psi^{\ast},\Psi]$ takes the form

\begin{equation}
\int_{0}^{\beta}d\tau\int
d^{2}\mathbf{r}\left[\Psi^{*}(\mathbf{r},\tau)(\partial_{\tau}-\mu+\Omega)\Psi(\mathbf{r},\tau)+\frac{1}{2}g|\Psi(\mathbf{r},\tau)|^{4}\right],\end{equation}
 in which $\beta=\frac{1}{k_{B}T}$ and $\mu$ is the chemical potential.
By variable rescaling :
$\frac{1}{mg\Omega}\left(\mu-\Omega\right)=a_{\mu},\tau=\frac{1}{mg\Omega}\tau^{\prime},\beta=\frac{1}{mg\Omega}\beta^{\prime},\mathbf{r}=\frac{1}{\sqrt{2m\Omega}}\mathbf{r}^{\prime},\Psi=\sqrt{2m\Omega}\Psi^{\prime},$
the partition function simplifies

\begin{equation}
\mathcal{Z}^{\prime}(\beta^{\prime},a_{\mu})=\int\mathcal{D}[\Psi^{\prime*}\Psi^{\prime}]\exp-\int_{0}^{\beta^{\prime}}d\tau^{\prime}\int
d^{2}\mathbf{r}^{\prime}\left[\Psi^{\prime*}(\mathbf{r}^{\prime},\tau^{\prime})(\partial_{\tau^{\prime}}-a_{\mu})\Psi^{\prime}(\mathbf{r}^{\prime},\tau^{\prime})+|\Psi^{\prime}(\mathbf{r}^{\prime},\tau^{\prime})|^{4}\right].\end{equation}
 With all primes omitted, we have

\begin{equation}
\mathcal{Z}(\beta,a_{\mu})=\int\mathcal{D}[\Psi^{\ast}\Psi]\exp-\int_{0}^{\beta}d\tau\int
d^{2}\mathbf{r}\left[\Psi^{\ast}(\mathbf{r},\tau)(\partial_{\tau}-a_{\mu})\Psi(\mathbf{r},\tau)+|\Psi(\mathbf{r},\tau)|^{4}\right].\label{eq:partition}\end{equation}
 Note that the kinetic part is absorbed in a shift of the effective
chemical potential $a_{\mu}$, and only the interaction part is
relevant. Hereafter we will work with the rescaled variables.

Using Landau gauge $\mathbf{A}=(-By,0)$ for the effective magnetic
field $\mathbf{B}=\frac{2m\Omega}{Q}\mathbf{e}_{z}$, we have the LLL
magnetic Bloch representation \cite{key-41,key-42}

\begin{equation}
\varphi_{\mathbf{k}}(\mathbf{r})=3^{\frac{1}{8}}\sum_{n=-\infty}^{\infty}\exp\left[-\frac{2\pi}{\sqrt{3}}\left(\frac{y}{d}-\frac{\sqrt{3}}{2}n-\frac{\sqrt{3}}{4\pi}k_{x}d\right)^{2}+i\left(\frac{\pi}{2}(n^{2}-n)+2\pi
n\frac{x}{d}+\frac{\sqrt{3}}{2}nk_{y}d+k_{x}x\right)\right].\end{equation}
 $\varphi(\mathbf{r})$, i.e. $\varphi_{\mathbf{k}}(\mathbf{r})$
with $\mathbf{k}=0$, is a superposition of the lowest Landau levels
of the kinetic part of the Hamiltonian and corresponds to an
Abrikosov triangular vortex-lattice with lattice spacing $d$. In the
rescaling length unit, $d$ equals $\sqrt{\frac{4\pi}{\sqrt{3}}}$ and
the primitive vectors of the vortex-lattice are
$\mathbf{d}_{1}=\sqrt{\frac{4\pi}{\sqrt{3}}}(1,0)$ ,
$\mathbf{d}_{2}=\sqrt{\frac{4\pi}{\sqrt{3}}}(\frac{1}{2},\frac{\sqrt{3}}{2})$.
$\varphi_{\mathbf{k}}(\mathbf{r})$ describes the transverse
oscillations of the vortex-lattice, and the primitive vectors of the
reciprocal lattice are
$\tilde{\mathbf{d}}_{1}=\sqrt{\frac{4\pi}{\sqrt{3}}}(\frac{\sqrt{3}}{2},-\frac{1}{2})$,
$\tilde{\mathbf{d}}_{2}=\sqrt{\frac{4\pi}{\sqrt{3}}}(0,1)$. The
vortex cores, i.e. the zero point of the function
$\varphi_{\mathbf{k}}(\mathbf{r})$ with
$\mathbf{k}=k_{1}\tilde{\mathbf{d}}_{1}+k_{2}\tilde{\mathbf{d}}_{2}$,
are uniformly distributed at sites
$(n_{1}-k_{2})\mathbf{d}_{1}+(n_{2}+\frac{1}{2}+k_{1})\mathbf{d}_{2}$
, where $n_{1}$and $n_{2}$ are integers. Note that both the lattice
cell and the Brillouin zone have an area of $2\pi$, and the vortex
number density is $n_{\upsilon}=\frac{1}{2\pi}$ in the rescaling
units.

\section{The Long-Range Correlations\label{sec:long-range correlation}}

As is known \cite{key-26}, for the ultrafast rotating two
dimensional Bose gas, there is no BEC in the thermodynamic limit
even at zero temperature. In spite of that, we will calculate
$U\left(1\right)$ invariant quantities, like the atom number
density, the free energy and the correlation functions, in the
perturbative framework, and show that the infrared divergences are
canceled, similar to the method used in two dimensional non-linear
$\sigma$ model \cite{key-46,key-47}. In two dimensional $O(N)$
non-linear $\sigma$ model, David in \cite{key-47} proved that, using
the {}``wrong'' spontaneously broken symmetry phase, any $O(N)$
invariant observable has an infrared finite weak coupling
perturbative expansion. In this paper, though we will only show some
$U\left(1\right)$ invariant quantities are also free of infrared
divergences at most to two loops in the perturbative framework, we
believe that it is true to all orders similar to the two dimensional
$O(N)$ non-linear $\sigma$ model. This method was extensively used
to study the vortex lattice in type II superconductors, for example,
in Ref. \cite{Rosenstein98,key-43}.

\bigskip{}

In this section, We calculate the condensate density perturbatively
and show it is zero in the thermodynamic limit. By calculating the
single-particle correlation function and the density fluctuation
correlation function, we obtain the algebraic decay exponent. The
atom number local density is also calculated, which shows that at
large filling fractions, the number density retains the
vortex-lattice configuration, while at small filling fractions,
quantum fluctuations tend to smooth away the vortex-lattice.

\subsection{The atom number density\label{sub:disprove BEC}}

In a usual fashion \cite{key-34,key-35,key-36,key-37}, we separate
the field as the condensate part and the fluctuation part

\begin{equation}
\Psi(\mathbf{r},\tau)=\sqrt{n_{c}}\varphi(\mathbf{r})+\psi(\mathbf{r},\tau),\end{equation}
where $n_{c}$, the condensate number density, is a real number
minimizing the free energy and the fluctuation part
$\psi(\mathbf{r},\tau)$ can be expanded as

\begin{equation}
\psi(\mathbf{r},\tau)=\frac{1}{\sqrt{A\beta}}\sum_{\mathbf{k}\in
BZ}\sum_{m}\psi_{\mathbf{k}m}\varphi_{\mathbf{k}}(\mathbf{r})e^{-\frac{i}{2}\theta_{\mathbf{k}}}e^{-i\omega_{m}\tau}\,.\end{equation}
 In powers of $\psi_{\mathbf{k}m}^{*}$ and $\psi_{\mathbf{k}m}$,
we divide the action $S$ into four parts

\begin{eqnarray}
S_{0} & = & \beta A\left(-a_{\mu}n_{c}+\beta_{A}n_{c}^{2}\right),\nonumber \\
S_{2} & = & \sum_{p}\left[\left(-i\omega_{m}-a_{\mu}+4n_{c}\beta_{\mathbf{k}}\right)\psi_{\mathbf{k}m}^{\ast}\psi_{\mathbf{k}m}+n_{c}|\gamma_{\mathbf{k}}|\left(\psi_{\mathbf{k}m}^{\ast}\psi_{-\mathbf{k}-m}^{\ast}+\psi_{\mathbf{k}m}\psi_{-\mathbf{k}-m}\right)\right],\nonumber \\
S_{3} & = & 2\sqrt{n_{c}}\frac{1}{\sqrt{A\beta}}\sum_{p_{1},p_{2},p_{3}}\left(\psi_{p_{1}}^{\ast}\psi_{p_{2}}^{\ast}\psi_{p_{3}}P_{p_{1}p_{2}p_{3}0}+c.c.\right),\\
S_{4} & = &
\frac{1}{A\beta}\sum_{p_{1},p_{2},p_{3},p_{4}}\psi_{p_{1}}^{\ast}\psi_{p_{2}}^{\ast}\psi_{p_{3}}\psi_{p_{4}}P_{p_{1}p_{2}p_{3}p_{4}},\nonumber
\end{eqnarray} where $p\equiv(\mathbf{k},\omega_{m})$ ,
$\psi_{p}\equiv\psi_{\mathbf{k}m}$ and
$\sum_{p}\equiv\sum_{\mathbf{k}\in BZ}\sum_{m}$ . The quadratic part
can be diagonalized as

\begin{equation}
S_{2}=\sum_{p}\left(-i\omega_{m}+\epsilon(\mathbf{k})\right)a_{\mathbf{k}m}^{*}a_{\mathbf{k}m}\,,\end{equation}
 by the following Bogoliubov transformation:

\begin{eqnarray}
a_{\mathbf{k}m} & = & u_{\mathbf{k}}\psi_{\mathbf{k}m}+\upsilon_{\mathbf{k}}\psi_{-\mathbf{k}-m}^{\ast},\label{eq:UniTrans1-1}\\
a_{\mathbf{k}m}^{\ast} & = &
u_{\mathbf{k}}\psi_{\mathbf{k}m}^{\ast}+\upsilon_{\mathbf{k}}\psi_{-\mathbf{k}-m},\label{eq:UniTrans2-1}\end{eqnarray}
and inversely,\begin{eqnarray}
\psi_{\mathbf{k}m} & = & u_{\mathbf{k}}a_{\mathbf{k}m}-\upsilon_{\mathbf{k}}a_{-\mathbf{k}-m}^{\ast},\label{eq:eq:UniTrans1-1}\\
\psi_{\mathbf{k}m}^{\ast} & = &
u_{\mathbf{k}}a_{\mathbf{k}m}^{\ast}-\upsilon_{\mathbf{k}}a_{-\mathbf{k}-m},\label{eq:eq:UniTrans2-1}\end{eqnarray}
where\begin{eqnarray}
u_{\mathbf{k}} & = & \sqrt{\frac{1}{2}\left(\frac{\epsilon_{0}(\mathbf{k})}{\epsilon(\mathbf{k})}+1\right)}\,,\nonumber \\
\upsilon_{\mathbf{k}} & = &
\sqrt{\frac{1}{2}\left(\frac{\epsilon_{0}(\mathbf{k})}{\epsilon(\mathbf{k})}-1\right)}\,,\end{eqnarray}
and\begin{eqnarray}
\epsilon_{0}(\mathbf{k}) & = & \left(-a_{\mu}+4n_{c}\beta_{\mathbf{k}}\right),\nonumber \\
\epsilon(\mathbf{k}) & = &
\sqrt{\left(-a_{\mu}+4n_{c}\beta_{\mathbf{k}}\right)^{2}-4n_{c}^{2}|\gamma_{\mathbf{k}}|^{2}}\,.\end{eqnarray}
Note that the free energy density $\mathcal{F}$ depends on two
parameters, $a_{\mu}$ and $n_{c}$, nevertheless, $n_{c}$ is related
to $a_{\mu}$ by the constraint
$\frac{\partial\mathcal{F}(n_{c},a_{\mu})}{\partial n_{c}}=0$ and
hence $n_{c}$ is renormalized order by order. In the zero-loop
order,

\begin{equation}
\mathcal{F}_{c}=-a_{\mu}n_{c}+\beta_{A}n_{c}^{2},\end{equation} and
the condensate density equals\begin{equation}
n_{c}^{(0)}=\frac{a_{\mu}}{2\beta_{A}},\end{equation} which is also
the total number density in the zero-loop order, denoted as $n_{0}$.
Now we want to calculate the total number density and the condensate
density to one-loop. The total number density $n$ is given
by\begin{equation} \frac{1}{A}\int d^{2}\mathbf{r}\left\langle
\Psi^{\dagger}(\mathbf{r})\Psi(\mathbf{r})\right\rangle
=n_{c}+\frac{1}{A\beta}\sum_{p}\left\langle
\psi_{\mathbf{k}m}^{\ast}\psi_{\mathbf{k}m}\right\rangle
\end{equation} where $n_{c}$ is the condensate density and
$\frac{1}{A\beta}\sum_{p}\left\langle
\psi_{\mathbf{k}m}^{\ast}\psi_{\mathbf{k}m}\right\rangle $ is the
density of the atoms outside the condensate. First we need to know
the one-loop correction to the condensate density, denoted as
$n_{c}^{(1)}$. Up to one-loop order, the free energy density takes
the form\begin{equation}
\mathcal{F}_{0+1}(a_{\mu},n_{c})=-a_{\mu}n_{c}+\beta_{A}n_{c}^{2}+\frac{1}{4\pi}\int_{BZ}\frac{d^{2}\mathbf{k}}{2\pi}\sqrt{\left(-a_{\mu}+4n_{c}\beta_{\mathbf{k}}\right)^{2}-4n_{c}^{2}|\gamma_{\mathbf{k}}|^{2}}\,.\end{equation}
Minimizing $\mathcal{F}_{0+1}$ with respect to $n_{c}$ leads to
\begin{equation}
n_{c}=\frac{a_{\mu}}{2\beta_{A}}-\frac{1}{2\pi\beta_{A}}\int_{BZ}\frac{d^{2}\mathbf{k}}{2\pi}\frac{\beta_{\mathbf{k}}\left(-a_{\mu}+4n_{c}\beta_{\mathbf{k}}\right)-n_{c}|\gamma_{\mathbf{k}}|^{2}}{\sqrt{\left(-a_{\mu}+4n_{c}\beta_{\mathbf{k}}\right)^{2}-4n_{c}^{2}|\gamma_{\mathbf{k}}|^{2}}}\,,\end{equation}
from which we see that the one-loop correction to $n_{c}$ equals
\begin{equation}
n_{c}^{(1)}=-\frac{1}{4\pi\beta_{A}}\int_{BZ}\frac{d^{2}\mathbf{k}}{2\pi}\frac{2\beta_{\mathbf{k}}\left(2\beta_{\mathbf{k}}-\beta_{A}\right)-|\gamma_{\mathbf{k}}|^{2}}{\sqrt{\left(2\beta_{\mathbf{k}}-\beta_{A}\right)^{2}-|\gamma_{\mathbf{k}}|^{2}}}\,.\end{equation}
The density of the atoms outside the condensate is given
by\begin{equation} \frac{1}{A\beta}\sum_{p}\left\langle
\psi_{\mathbf{k}m}^{\ast}\psi_{\mathbf{k}m}\right\rangle
_{1-loop}=\frac{1}{4\pi}\int_{BZ}\frac{d^{2}\mathbf{k}}{2\pi}\frac{E_{0}(\mathbf{k})}{E(\mathbf{k})}\,,\end{equation}
where $E_{0}(\mathbf{k})$ and $E(\mathbf{k})$ are defined as
\begin{eqnarray}
E_{0}(\mathbf{k}) & = & 2\beta_{\mathbf{k}}-\beta_{A}\,,\label{eq:E0}\\
E(\mathbf{k}) & = &
\sqrt{\left(2\beta_{\mathbf{k}}-\beta_{A}\right)^{2}-|\gamma_{\mathbf{k}}|^{2}}\,.\label{eq:E}\end{eqnarray}
The total number density, $n_{0+1}$, is equal to
$n_{c}^{(0)}+n_{c}^{(1)}+\frac{1}{A\beta}\sum_{p}\left\langle
\psi_{\mathbf{k}m}^{\ast}\psi_{\mathbf{k}m}\right\rangle _{1-loop}$
. The filling fraction, $n/n_{\upsilon}$, is given by $2\pi n_{0+1}$
at the one-loop level.

Obviously, $n_{c}^{(1)}$ and $\frac{1}{A\beta}\sum_{p}\left\langle
\psi_{\mathbf{k}m}^{\ast}\psi_{\mathbf{k}m}\right\rangle _{1-loop}$
both contain infrared divergences, but in the sum the divergences
are canceled and $n_{c}^{(1)}+\frac{1}{A\beta}\sum_{p}\left\langle
\psi_{\mathbf{k}m}^{\ast}\psi_{\mathbf{k}m}\right\rangle _{1-loop}$
, which is the one-loop correction to the total number density,
equals

\begin{equation}
n_{1}=-\frac{1}{4\pi\beta_{A}}\int_{BZ}\frac{d^{2}\mathbf{k}}{2\pi}E(\mathbf{k})\simeq-0.023.\label{eq:n1-1}\end{equation}
The condensate density, $n_{c}=n_{c}^{(0)}+n_{c}^{(1)}$, is infrared
divergent, $n_{c}^{(0)}+n_{c}^{(1)}=c\left(1-\alpha\ln L\right)$,
where $c$ and $\alpha$ are positive constants and $L$ is the system
size. If we can calculate $n_{c}$ to all loops, $n_{c}\simeq
c\left(1-\alpha\ln L+\frac{1}{2}(\alpha\ln
L)^{2}+\cdots\right)\simeq c\exp\left(-\alpha\ln
L\right)=cL^{-\alpha}$. When we take $L\rightarrow\infty$,
$n_{c}\rightarrow0$. Therefore, in the thermodynamic limit, there
will be no condensate, as is also shown by J. Sinova etc.
\cite{key-26}. For a finite system, there is a finite infrared
cutoff $\sim\frac{1}{L}$, and the condensate density,
$n_{c}=cL^{-\alpha}$ , will be finite.

Up to one-loop, the local density $n(\mathbf{r})$ is given by
\begin{equation} \left\langle
\Psi^{\dagger}(\mathbf{r})\Psi(\mathbf{r})\right\rangle
=(n_{c}^{(0)}+n_{c}^{(1)})\varphi^{\ast}(\mathbf{r})\varphi(\mathbf{r})+\frac{1}{4\pi}\int_{BZ}\frac{d^{2}\mathbf{k}}{2\pi}\frac{E_{0}(\mathbf{k})}{E(\mathbf{k})}\varphi_{\mathbf{k}}^{\ast}(\mathbf{r})\varphi_{\mathbf{k}}(\mathbf{r})\,,\label{eq:local_density}\end{equation}
where $n_{c}^{(0)}\varphi^{\ast}(\mathbf{r})\varphi(\mathbf{r})$ is
the mean field local density, denoted as $n_{0}(\mathbf{r})$.
Obviously, in Eq. \eqref{eq:local_density} $n_{c}^{(1)}$and the last
term both contain infrared divergences, nevertheless, by arranging
them properly we see the divergences are canceled and the one-loop
correction to the total local density is obtained as
\begin{eqnarray}
n_{1}(\mathbf{r}) & = & n_{c}^{(1)}\varphi^{\ast}(\mathbf{r})\varphi(\mathbf{r})+\frac{1}{4\pi}\int_{BZ}\frac{d^{2}\mathbf{k}}{2\pi}\frac{E_{0}(\mathbf{k})}{E(\mathbf{k})}\varphi_{\mathbf{k}}^{\ast}(\mathbf{r})\varphi_{\mathbf{k}}(\mathbf{r})\nonumber \\
 & = & n_{1}\varphi^{\ast}(\mathbf{r})\varphi(\mathbf{r})+\frac{1}{4\pi}\int_{BZ}\frac{d^{2}\mathbf{k}}{2\pi}\frac{E_{0}(\mathbf{k})}{E(\mathbf{k})}\left(\varphi_{\mathbf{k}}^{\ast}(\mathbf{r})\varphi_{\mathbf{k}}(\mathbf{r})-\varphi^{\ast}(\mathbf{r})\varphi(\mathbf{r})\right),\end{eqnarray}
in which the second term is free of divergences.

\begin{figure}
\includegraphics[scale=1.5]{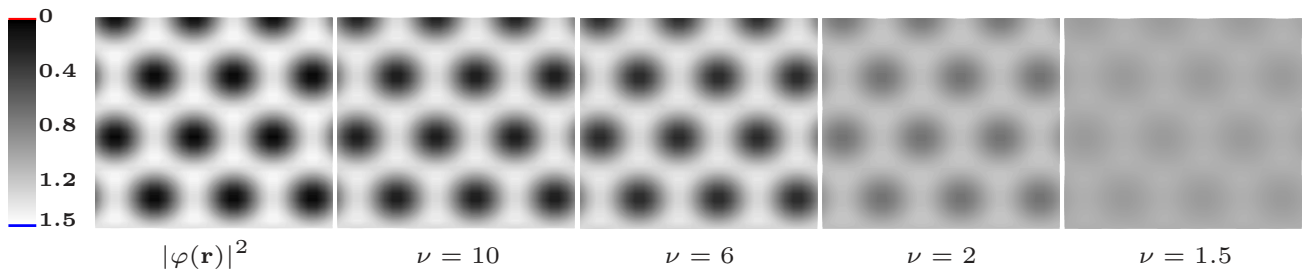}
\caption{The mean field atom number density distribution forms a
perfect triangular lattice, described by the function
$|\varphi(\mathbf{r})|^{2}$. Quantum fluctuations tend to smooth the
mean field vortex-lattice. As the filling fraction lowers, quantum
fluctuations increase and the vortex-lattice becomes more smooth.}

\label{fig:local density}
\end{figure}

Note that the mean field local density,
$n_{0}(\mathbf{r})=n_{c}^{(0)}\varphi^{*}(\mathbf{r})\varphi(\mathbf{r})$
, forms a vortex-lattice, and the one-loop correction,
$n_{1}(\mathbf{r})$ , includes quantum fluctuations. In order to
explore quantum fluctuation effects on the atom number density
configuration, we plot the mean field density distribution,
$\frac{n_{0}(\mathbf{r})}{n_{0}}=\varphi^{*}(\mathbf{r})\varphi(\mathbf{r})$,
and the total density distribution,
$\frac{n_{0}(\mathbf{r})+n_{1}(\mathbf{r})}{n_{0}+n_{1}}$, at
different filling fractions. From Fig. \ref{fig:local density}, one
finds that quantum fluctuations tend to smooth the vortex-lattice.
At large filling fractions, the number density still retains the
vortex-lattice configuration, while at small filling fractions, the
vortex-lattice
is smoothed away. %

\subsection{The single-particle correlation function}

Up to one-loop, the single-particle correlation function
$\left\langle
\Psi^{\dagger}(\mathbf{r}_{1})\Psi(\mathbf{r}_{2})\right\rangle $ is
equal to

\begin{eqnarray}
\left\langle \Psi^{\dagger}(\mathbf{r}_{1})\Psi(\mathbf{r}_{2})\right\rangle  & = & \left(n_{c}^{(0)}+n_{c}^{(1)}\right)\varphi^{\ast}(\mathbf{r}_{1})\varphi(\mathbf{r}_{2})+\left\langle \psi^{\dagger}(\mathbf{r}_{1})\psi(\mathbf{r}_{2})\right\rangle \nonumber \\
 & = & \left(n_{c}^{(0)}+\lambda n_{c}^{(1)}\right)\varphi^{\ast}(\mathbf{r}_{1})\varphi(\mathbf{r}_{2})+\lambda\left\langle \psi^{\dagger}(\mathbf{r}_{1})\psi(\mathbf{r}_{2})\right\rangle ,\label{eq:correlation}\end{eqnarray}
 where $\lambda$ is used to keep trace of the loop order and should
be set to $1$ in the end. Similar to Eq. \eqref{eq:local_density},
the above equation can be arranged as

\begin{eqnarray}
 &  & \left(n_{c}^{(0)}+\lambda n_{c}^{(1)}+\lambda\frac{1}{4\pi}\int_{BZ}\frac{d^{2}\mathbf{k}}{2\pi}\frac{E_{0}(\mathbf{k})}{E(\mathbf{k})}\right)\varphi^{\ast}(\mathbf{r}_{1})\varphi(\mathbf{r}_{2})+\lambda f(\mathbf{r}_{1},\mathbf{r}_{2})\nonumber \\
 & = & \left(n_{c}^{(0)}+\lambda n_{1}\right)\varphi^{\ast}(\mathbf{r}_{1})\varphi(\mathbf{r}_{2})+\lambda f(\mathbf{r}_{1},\mathbf{r}_{2})\,,\end{eqnarray}
 where $f(\mathbf{r}_{1},\mathbf{r}_{2})$ is given by$\frac{1}{4\pi}\int_{BZ}\frac{d^{2}\mathbf{k}}{2\pi}\frac{E_{0}(\mathbf{k})}{E(\mathbf{k})}\left[\varphi_{\mathbf{k}}^{\ast}(\mathbf{r}_{1})\varphi_{\mathbf{k}}(\mathbf{r}_{2})-\varphi^{*}(\mathbf{r}_{1})\varphi(\mathbf{r}_{2})\right]$
and free of divergences.

Now we define the relative coordinate
$\mathbf{r}\equiv\mathbf{r}_{1}-\mathbf{r}_{2}$, and the center
coordinate
$\mathbf{R}\equiv\frac{\mathbf{r}_{1}+\mathbf{r}_{2}}{2}$. We take
$\mathbf{R=0}$ for simplicity and analyze the large $|\mathbf{r}|$
limit. Based on numerical calculations, we find

\begin{equation}
\lim_{\mathbf{|\mathbf{r}|\rightarrow\infty}}f(\mathbf{r}_{1},\mathbf{r}_{2})\simeq-0.18\ln|\mathbf{r}|\varphi^{\ast}(\mathbf{r}_{1})\varphi(\mathbf{r}_{2})\end{equation}
Therefore\begin{eqnarray}
\lim_{\mathbf{|\mathbf{r}|\rightarrow\infty}}\left\langle \Psi^{\dagger}(\mathbf{r}_{1})\Psi(\mathbf{r}_{2})\right\rangle  & \simeq & \left(n_{c}^{(0)}+\lambda n_{1}\right)\varphi^{\ast}(\mathbf{r}_{1})\varphi(\mathbf{r}_{2})-0.18\lambda\ln|\mathbf{r}|\varphi^{\ast}(\mathbf{r}_{1})\varphi(\mathbf{r}_{2})\nonumber \\
 & = & \left(n_{c}^{(0)}+\lambda n_{1}\right)\varphi^{\ast}(\mathbf{r}_{1})\varphi(\mathbf{r}_{2})\left(1-\alpha\ln|\mathbf{r}|\right)\end{eqnarray}
where $\alpha=\frac{0.18\lambda}{n_{c}^{(0)}+\lambda n_{1}}$, and to
one loop order
$\alpha\approx\frac{0.18\lambda}{n_{c}^{(0)}}=\frac{0.18}{n_{c}^{(0)}}$.
Of course this result is not physical if we only include one-loop
correction, as $\lim_{\mathbf{r\rightarrow\infty}}\left\langle
\Psi^{\dagger}(\mathbf{r}_{1}))\Psi(\mathbf{r}_{2})\right\rangle
\rightarrow-\infty$ . We shall include all order in perturbation
theory to get the physical result. We argue, in all order, similar
to the calculation in Ref. \cite{key-43},

\begin{eqnarray}
\lim_{\mathbf{r\rightarrow\infty}}\left\langle \Psi^{\dagger}(\mathbf{r}_{1})\Psi(\mathbf{r}_{2})\right\rangle  & \propto & \varphi^{\ast}(\mathbf{r}_{1})\varphi(\mathbf{r}_{2})\left(1-\alpha^{\prime}\ln|\mathbf{r}|+\frac{1}{2}(\alpha^{\prime}\ln|\mathbf{r}|)^{2}+...\right)\label{eq:KT}\\
 & = & \varphi^{\ast}(\mathbf{r}_{1})\varphi(\mathbf{r}_{2})\left\vert \mathbf{r}\right\vert ^{-\alpha^{\prime}}\nonumber \end{eqnarray}
To one loop, $\alpha^{\prime}=\alpha=\frac{0.18}{n_{c}^{(0)}}$.
similar calculations can be done in usual BKT (Berezinsky,
Kosterlitz and Thouless) phase transition systems, and the algebraic
decay exponent can be obtained correctly \cite{key-44,key-45}.

One will wonder in usual BKT phase transition systems, the phase
transition is continuous. In the vortex lattice phase, from Eq.
\eqref{eq:KT}, the correlation decays algebraically, and the
rotational symmetry is broken as the factor
$\varphi^{\ast}(\mathbf{r}_{1})\varphi(\mathbf{r}_{2})$ in Eq.
\eqref{eq:KT} is not rotationally invariant. In the vortex liquid
phase, the correlation decays exponentially, and the rotational
symmetry is unbroken. Therefore, the phase transition is the
spontaneous breaking of the rotational symmetry. As in usual solid
to liquid phase transition, the phase transition is a first order
melting transition.

\subsection{The density fluctuation correlation function}

In the following, we will calculate the density fluctuation
correlation to one-loop,\begin{eqnarray}
\left\langle \delta\hat{n}(\mathbf{r}_{1})\delta\hat{n}(\mathbf{r}_{2})\right\rangle  & = & \left\langle \left[\hat{n}(\mathbf{r}_{1})-\left\langle \hat{n}(\mathbf{r}_{1})\right\rangle \right]\left[\hat{n}(\mathbf{r}_{2})-\left\langle \hat{n}(\mathbf{r}_{2})\right\rangle \right]\right\rangle \nonumber \\
 & = & \left\langle \hat{n}(\mathbf{r}_{1})\hat{n}(\mathbf{r}_{2})\right\rangle -\left\langle \hat{n}(\mathbf{r}_{1})\right\rangle \left\langle \hat{n}(\mathbf{r}_{2})\right\rangle .\end{eqnarray}
 As shown in subsection \ref{sub:disprove BEC}, there is no condensate
and thus the system retains the $U(1)$ symmetry, therefore, only the
$U(1)$ gauge invariant terms in the contractions remain

\begin{eqnarray}
 &  & \left\langle \hat{n}(\mathbf{r}_{1})\hat{n}(\mathbf{r}_{2})\right\rangle -\left\langle \hat{n}(\mathbf{r}_{1})\right\rangle \left\langle \hat{n}(\mathbf{r}_{2})\right\rangle \nonumber \\
 & = & \left\langle \Psi^{\dagger}(\mathbf{r}_{1})\Psi(\mathbf{r}_{2})\right\rangle \left\langle \Psi(\mathbf{r}_{1})\Psi^{\dagger}(\mathbf{r}_{2})\right\rangle +\left\langle \Psi^{\dagger}(\mathbf{r}_{1})\Psi(\mathbf{r}_{1})\Psi^{\dagger}(\mathbf{r}_{2})\Psi(\mathbf{r}_{2})\right\rangle _{c}\,.\label{eq:density-fluctuation}\end{eqnarray}
The term$\left\langle
\Psi^{\dagger}(\mathbf{r}_{1})\Psi(\mathbf{r}_{2})\right\rangle $ is
investigated in the last subsection. In the
large$|\mathbf{r}_{1}-\mathbf{r}_{2}|$ limit, the first term in Eq.
\eqref{eq:density-fluctuation} falls off as an inverse power of the
separation distance $|\mathbf{r}_{1}-\mathbf{r}_{2}|$ ,
\begin{equation}
\lim_{|\mathbf{r}_{1}-\mathbf{r}_{2}|\rightarrow\infty}\left\langle
\Psi^{\dagger}(\mathbf{r}_{1})\Psi(\mathbf{r}_{2})\right\rangle
\left\langle
\Psi(\mathbf{r}_{1})\Psi^{\dagger}(\mathbf{r}_{2})\right\rangle
\sim\left\vert \varphi(\mathbf{r}_{1})\right\vert ^{2}\left\vert
\varphi(\mathbf{r}_{2})\right\vert
^{2}|\mathbf{r}_{1}-\mathbf{r}_{2}|^{-2\alpha}.\end{equation} The
second term in Eq. \eqref{eq:density-fluctuation} is the two-body
connected Green's function and it is hard to calculate
non-perturbatively. But we speculate that this term will not alter
the asymptotic behavior of $\left\langle
\delta\hat{n}(\mathbf{r}_{1})\delta\hat{n}(\mathbf{r}_{2})\right\rangle
$ in the large$|\mathbf{r}_{1}-\mathbf{r}_{2}|$ limit, and hence the
density fluctuation correlation function also decays algebraically
with the distance in the large distance limit. The result will
indicate, near the Brag peak, the structure function
$S\left(\boldsymbol{Q+k}\right)$ where $\boldsymbol{Q}$ belongs to
the reciprocal lattice and $k$ is small will have a scaling
$\lim_{k\rightarrow0}S\left(\boldsymbol{Q+k}\right)\propto\left\vert
\boldsymbol{k}\right\vert ^{-(2-2\alpha)}$.

Similar calculation can be done for vortex lattice in type II
superconductors and the density fluctuation correlation will be
shown to have similar behavior. For the vortex lattice in type II
superconductors near the Brag peak, the structure function
$S\left(\boldsymbol{Q+k}\right)$ will have a scaling
$\lim_{k\rightarrow0}S\left(\boldsymbol{Q+k}\right)\propto\left\vert
\boldsymbol{k}\right\vert ^{-\eta}$ (details will be published
elsewhere).

\section{The Free Energy Density Calculation By Loop Expansion \label{perturbation theory}}

In this section, we shall calculate the free energy density by loop
expansion up to two-loop and show the cancelation of infrared
divergences.

\subsection{Mean-field contribution}

In the saddle point approximation,

\begin{equation}
\frac{\delta S[\Psi^{\ast},\Psi]}{\delta\Psi^{\ast}}=0,\frac{\delta
S[\Psi^{\ast},\Psi]}{\delta\Psi}=0,\end{equation} with the LLL
constraint, one obtains \cite{key-3,key-29}\begin{equation}
\Psi_{0}(\mathbf{r})=\sqrt{\frac{a_{\mu}}{2\beta_{A}}}\varphi(\mathbf{r}),\end{equation}
where $\varphi(\mathbf{r})=\varphi_{\mathbf{k}=0}(\mathbf{r})$ ,
$\beta_{A}=\frac{1}{2\pi}\int_{cell}d^{2}\mathbf{r}|\varphi(\mathbf{r})|^{4}$.
Obviously, the saddle point approximation is equivalent to the mean
field GP equation. We have the mean field contribution to the free
energy density \begin{equation}
\mathcal{F}_{0}(a_{\mu})=-\frac{a_{\mu}^{2}}{4\beta_{A}},\end{equation}
and the mean field contribution to the number
density\begin{equation}
n_{0}(a_{\mu})=-\frac{\partial\mathcal{F}_{c}(a_{\mu})}{\partial
a_{\mu}}=\frac{a_{\mu}}{2\beta_{A}}.\end{equation}

\subsection{One-loop correction}

Following the loop expansion procedure presented in \cite{key-31},
we set \begin{equation}
\Psi(\mathbf{r},\tau)=\sqrt{\frac{a_{\mu}}{2\beta_{A}}}\varphi(\mathbf{r})+\psi(\mathbf{r},\tau)\,,\end{equation}
 in which $\sqrt{\frac{a_{\mu}}{2\beta_{A}}}\varphi(\mathbf{r})$
is the mean field part and $\psi(\mathbf{r},\tau)$ is the higher
order corrections, and then expand $S[\Psi^{*},\Psi]$ in powers of
$\psi^{*}(\mathbf{r},\tau)$ and $\psi(\mathbf{r},\tau)$. In the
magnetic Bloch representation, $\psi(\mathbf{r},\tau)$ can be
expanded as

\begin{equation}
\psi(\mathbf{r},\tau)=\frac{1}{\sqrt{A\beta}}\sum_{\mathbf{k}\in
BZ}\sum_{m}\psi_{\mathbf{k}m}\varphi_{\mathbf{k}}(\mathbf{r})e^{-\frac{i}{2}\theta_{\mathbf{k}}}e^{-i\omega_{m}\tau}\end{equation}
where $A$ is the area of the sample, $\omega_{m}=\frac{2\pi
m}{\beta}$ is the bosonic Matsubara frequency, and
$\theta_{\mathbf{k}}$ is defined in Eq. \eqref{eq:theta_k}. In
powers of $\psi_{\mathbf{k}m}^{\ast}$ and $\psi_{\mathbf{k}m}$, we
divide the action into four parts\begin{eqnarray}
S_{0} & = & A\beta\left(-\frac{a_{\mu}^{2}}{4\beta_{A}}\right),\nonumber \\
S_{2} & = & \sum_{p}[\left(-i\omega_{m}+\frac{a_{\mu}}{\beta_{A}}\left(2\beta_{\mathbf{k}}-\beta_{A}\right)\right)\psi_{\mathbf{k}m}^{\ast}\psi_{\mathbf{k}m}+\frac{1}{2}\frac{a_{\mu}}{\beta_{A}}|\gamma_{\mathbf{k}}|\left(\psi_{\mathbf{k}m}^{\ast}\psi_{-\mathbf{k}-m}^{\ast}+\psi_{\mathbf{k}m}\psi_{-\mathbf{k}-m}\right)],\nonumber \\
S_{3} & = & \sqrt{\frac{2a_{\mu}}{\beta_{A}A\beta}}\sum_{p_{1},p_{2},p_{3}}\left(\psi_{p_{1}}^{\ast}\psi_{p_{2}}^{\ast}\psi_{p_{3}}P_{p_{1}p_{2}p_{3}0}+c.c.\right),\\
S_{4} & = &
\frac{1}{A\beta}\sum_{p_{1},p_{2},p_{3},p_{4}}\psi_{p_{1}}^{\ast}\psi_{p_{2}}^{\ast}\psi_{p_{3}}\psi_{p_{4}}P_{p_{1}p_{2}p_{3}p_{4}},\nonumber
\end{eqnarray} where $p\equiv(\mathbf{k},\omega_{m})$ ,
$\psi_{p}\equiv\psi_{\mathbf{k}m}$ ,
$\sum_{p}\equiv\sum_{\mathbf{k}\in BZ}\sum_{m}$ and \begin{eqnarray}
\beta_{\mathbf{k}} & = & \frac{1}{2\pi}\int_{cell}d^{2}\mathbf{r}\varphi_{\mathbf{k}}^{\ast}(\mathbf{r})\varphi^{\ast}(\mathbf{r})\varphi_{\mathbf{k}}(\mathbf{r})\varphi(\mathbf{r}),\nonumber \\
\gamma_{\mathbf{k}} & = & \frac{1}{2\pi}\int_{cell}d^{2}\mathbf{r}\varphi^{\ast}(\mathbf{r})\varphi^{\ast}(\mathbf{r})\varphi_{\mathbf{k}}(\mathbf{r})\varphi_{\mathbf{-k}}(\mathbf{r}),\nonumber \\
e^{i\theta_{\mathbf{k}}} & = & \frac{\gamma_{\mathbf{k}}}{|\gamma_{\mathbf{k}}|},\label{eq:theta_k}\\
P_{p_{1}p_{2}p_{3}p_{4}} & = &
\delta_{m_{1}+m_{2},m_{3}+m_{4}}\int_{cell}\frac{d^{2}\mathbf{r}}{2\pi}\varphi_{\mathbf{k}_{1}}^{\ast}(\mathbf{r})\varphi_{\mathbf{k}_{2}}^{\ast}(\mathbf{r})\varphi_{\mathbf{k}_{3}}(\mathbf{r})\varphi_{\mathbf{k}_{4}}(\mathbf{r})e^{\frac{i}{2}(\theta_{\mathbf{k}_{1}}+\theta_{\mathbf{k}_{2}}-\theta_{\mathbf{k}_{3}}-\theta_{\mathbf{k}_{4}})}.\nonumber
\end{eqnarray} In the one-loop approximation, we keep only the
quadratic part of the action and diagonalize it as

\begin{equation}
S_{2}=\sum_{p}\left(-i\omega_{m}+\epsilon(\mathbf{k})\right)a_{\mathbf{k}m}^{\ast}a_{\mathbf{k}m}\,,\end{equation}
where\begin{equation}
\epsilon(\mathbf{k})=\frac{a_{\mu}}{\beta_{A}}\sqrt{\left(2\beta_{\mathbf{k}}-\beta_{A}\right)^{2}-|\gamma_{\mathbf{k}}|^{2}}\,,\end{equation}
by the following Bogoliubov transformation:

\begin{eqnarray}
a_{\mathbf{k}m} & = & u_{\mathbf{k}}\psi_{\mathbf{k}m}+\upsilon_{\mathbf{k}}\psi_{-\mathbf{k}-m}^{\ast}\,,\label{eq:UniTrans1}\\
a_{\mathbf{k}m}^{\ast} & = &
u_{\mathbf{k}}\psi_{\mathbf{k}m}^{\ast}+\upsilon_{\mathbf{k}}\psi_{-\mathbf{k}-m}\,,\label{eq:UniTrans2}\end{eqnarray}
 and inversely,\begin{eqnarray}
\psi_{\mathbf{k}m} & = & u_{\mathbf{k}}a_{\mathbf{k}m}-\upsilon_{\mathbf{k}}a_{-\mathbf{k}-m}^{*},\label{eq:eq:UniTrans1}\\
\psi_{\mathbf{k}m}^{*} & = &
u_{\mathbf{k}}a_{\mathbf{k}m}^{*}-\upsilon_{\mathbf{k}}a_{-\mathbf{k}-m},\label{eq:eq:UniTrans2}\end{eqnarray}
 where $u_{\mathbf{k}}=\sqrt{\frac{1}{2}\left(\frac{E_{0}(\mathbf{k})}{E(\mathbf{k})}+1\right)}$,$\upsilon_{\mathbf{k}}=\sqrt{\frac{1}{2}\left(\frac{E_{0}(\mathbf{k})}{E(\mathbf{k})}-1\right)}$,
$E_{0}(\mathbf{k})$ and $E(\mathbf{k})$ are defined in Eqs.
\eqref{eq:E0} and \eqref{eq:E}. By Taylor expanding
$\beta_{\mathbf{k}}$ and $|\gamma_{\mathbf{k}}|$ , we find the
excitation $\epsilon(\mathbf{k})$ has a quadratic dispersion at
small wave vectors, i.e.
$\lim_{\mathbf{k}\rightarrow0}\epsilon(\mathbf{k})\sim k^{2}$, which
is consistent with previous results \cite{key-26,key-32,key-33}. The
one-loop contribution to the free energy
density,$\mathcal{F}_{1}(a_{\mu})$, takes the form \begin{eqnarray}
 &  & -\frac{1}{A}\frac{1}{\beta}\ln\int\mathcal{D}[a^{\ast}a]\exp-\sum_{p}\left(-i\omega_{m}+\epsilon(\mathbf{k})\right)a_{\mathbf{k}m}^{\ast}a_{\mathbf{k}m}\nonumber \\
 & = & \frac{1}{A}\sum_{\mathbf{k}\in BZ}\left[\frac{1}{2}\epsilon(\mathbf{k})+\frac{1}{\beta}\ln\left(1-e^{-\beta\epsilon_{\mathbf{k}}}\right)\right].\end{eqnarray}
 By setting the area $A$ to infinity and the temperature $T$ to
zero, we obtain \begin{equation}
\mathcal{F}_{1}(a_{\mu})=\frac{a_{\mu}}{4\pi\beta_{A}}\left\langle
E(\mathbf{k})\right\rangle _{\mathbf{k}},\end{equation}
 where $\left\langle \cdots\right\rangle _{\mathbf{k}}\equiv\int_{BZ}\frac{d^{2}\mathbf{k}}{2\pi}$,
means average over the Brillouin zone. The one-loop correction to
the number density, $n_{1}(a_{\mu})$, is equal to

\begin{equation}
-\frac{\partial\mathcal{F}_{1}(a_{\mu})}{\partial
a_{\mu}}=-\frac{1}{4\pi\beta_{A}}\left\langle
E(\mathbf{k})\right\rangle _{\mathbf{k}},\label{eq:n1}\end{equation}
which is consistent with the result obtained in the last section as
in Eq.  \eqref{eq:n1-1}.

\subsection{Two-loop correction}

By the Bogoliubov transformation shown in Eqs.
\eqref{eq:eq:UniTrans1} and \eqref{eq:eq:UniTrans2}, we switch to
the field $a_{\mathbf{k}m}^{\ast}$, $a_{\mathbf{k}m}$, and write the
cubic and quartic part as

\begin{equation}
S_{3}=\sqrt{\frac{2a_{\mu}}{\beta_{A}}}\frac{1}{\sqrt{A\beta}}\sum_{p_{1},p_{2},p_{3}}[a_{p_{1}}a_{p_{2}}a_{p_{3}}^{\ast}(\Lambda_{p_{1}p_{2}p_{3}}-\Lambda_{p_{1}p_{2}p_{3}}^{\prime})+a_{p_{1}}a_{p_{2}}a_{p_{3}}({\textstyle
\prod_{p_{1}p_{2}p_{3}}^{\prime}}-{\textstyle
\prod_{p_{1}p_{2}p_{3}}})]+c.c.\end{equation}

and \begin{eqnarray}
S_{4} & = & \{\frac{1}{A\beta}\sum_{p_{1},p_{2},p_{3},p_{4}}[a_{p_{1}}a_{p_{2}}a_{p_{3}}a_{p_{4}}P_{p_{1}p_{2}-p_{3}-p_{4}}\upsilon_{\mathbf{k}_{1}}\upsilon_{\mathbf{k}_{2}}u_{\mathbf{k}_{3}}u_{\mathbf{k}_{4}}\nonumber \\
 &  & -2a_{p_{1}}^{\ast}a_{p_{2}}a_{p_{3}}a_{p_{4}}(P_{p_{1}-p_{4}p_{2}p_{3}}u_{\mathbf{k}_{1}}u_{\mathbf{k}_{2}}u_{\mathbf{k}_{3}}\upsilon_{\mathbf{k}_{4}}+P_{p_{2}p_{3}p_{1}-p_{4}}\upsilon_{\mathbf{k}_{1}}\upsilon_{\mathbf{k}_{2}}\upsilon_{\mathbf{k}_{3}}u_{\mathbf{k}_{4}})]+c.c.\}\nonumber \\
 &  & +\frac{1}{A\beta}\sum_{p_{1},p_{2},p_{3},p_{4}}a_{p_{1}}^{\ast}a_{p_{2}}^{\ast}a_{p_{3}}a_{p_{4}}[4P_{p_{1}-p_{4}p_{3}-p_{2}}u_{\mathbf{k}_{1}}u_{\mathbf{k}_{3}}\upsilon_{\mathbf{k}_{2}}\upsilon_{\mathbf{k}_{4}}\\
 &  & +P_{p_{1}p_{2}p_{3}p_{4}}u_{\mathbf{k}_{1}}u_{\mathbf{k}_{2}}u_{\mathbf{k}_{3}}u_{\mathbf{k}_{4}}+P_{p_{3}p_{4}p_{1}p_{2}}\upsilon_{\mathbf{k}_{1}}\upsilon_{\mathbf{k}_{2}}\upsilon_{\mathbf{k}_{3}}\upsilon_{\mathbf{k}_{4}}]\,,\nonumber \end{eqnarray}
where $a_{p}\equiv a_{\mathbf{k}m}$ and

\begin{align}
{\textstyle \prod_{p_{1}p_{2}p_{3}}}= & \frac{1}{3}\left(P_{0-p_{1}p_{2}p_{3}}u_{\mathbf{k}_{2}}u_{\mathbf{k}_{3}}\upsilon_{\mathbf{k}_{1}}+P_{0-p_{2}p_{3}p_{1}}u_{\mathbf{k}_{1}}u_{\mathbf{k}_{3}}\upsilon_{\mathbf{k}_{2}}+P_{0-p_{3}p_{2}p_{1}}u_{\mathbf{k}_{1}}u_{\mathbf{k}_{2}}\upsilon_{\mathbf{k}_{3}}\right),\nonumber \\
{\textstyle \prod_{p_{1}p_{2}p_{3}}^{\prime}}= &
\frac{1}{3}\left(P_{p_{3}p_{2}-p_{1}0}\upsilon_{\mathbf{k}_{2}}\upsilon_{\mathbf{k}_{3}}u_{\mathbf{k}_{1}}+P_{p_{1}p_{3}-p_{2}0}\upsilon_{\mathbf{k}_{1}}\upsilon_{\mathbf{k}_{3}}u_{\mathbf{k}_{2}}+P_{p_{1}p_{2}-p_{3}0}\upsilon_{\mathbf{k}_{1}}\upsilon_{\mathbf{k}_{2}}u_{\mathbf{k}_{3}}\right),\end{align}

and\begin{eqnarray}
\Lambda_{p_{1}p_{2}p_{3}} & = & P_{0p_{1}p_{3}-p_{2}}\upsilon_{\mathbf{k}_{1}}\upsilon_{\mathbf{k}_{3}}u_{\mathbf{k}_{2}}+P_{0p_{2}p_{3}-p_{1}}\upsilon_{\mathbf{k}_{2}}\upsilon_{\mathbf{k}_{3}}u_{\mathbf{k}_{1}}+P_{0p_{3}p_{2}p_{1}}u_{\mathbf{k}_{1}}u_{\mathbf{k}_{2}}u_{\mathbf{k}_{3}},\nonumber \\
\Lambda_{p_{1}p_{2}p_{3}}^{\prime} & = &
P_{-p_{2}p_{3}p_{1}0}u_{\mathbf{k}_{1}}u_{\mathbf{k}_{3}}\upsilon_{\mathbf{k}_{2}}+P_{-p_{1}p_{3}p_{2}0}u_{\mathbf{k}_{2}}u_{\mathbf{k}_{3}}\upsilon_{\mathbf{k}_{1}}+P_{p_{1}p_{2}p_{3}0}\upsilon_{\mathbf{k}_{1}}\upsilon_{\mathbf{k}_{2}}\upsilon_{\mathbf{k}_{3}}.\end{eqnarray}

\begin{figure}
\includegraphics[scale=0.7]{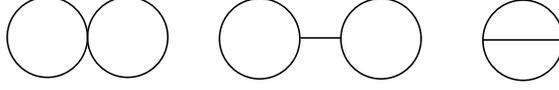}
\caption{The two-loop Feynman diagrams.}
\label{fig:The-two-loop-Feynman}
\end{figure}

The two-loop contribution to the free energy density,
$\mathcal{F}_{2}(a_{\mu})$, takes the form

\begin{equation}
-\frac{1}{\beta
A}\left[\ln\frac{\int\mathcal{D}[a^{\ast},a]\exp-\left(S_{2}+S_{3}+S_{4}\right)}{\int\mathcal{D}[a^{\ast},a]\exp-S_{2}}\right]_{2-loop}=\frac{1}{\beta
A}\left(\left\langle S_{4}\right\rangle -\frac{1}{2}\left\langle
S_{3}S_{3}\right\rangle \right),\end{equation} where $\left\langle
\cdots\right\rangle $denotes the sum of all the connected Feynman
diagrams with $G_{p}=\frac{1}{-i\omega_{m}+\epsilon(\mathbf{k})}$ as
a propagator. The two-loop Feynman diagrams are depicted in Fig.
\ref{fig:The-two-loop-Feynman}. The contribution from the diagram
{}\textquotedblleft $\infty $\textquotedblright\ equals

\begin{eqnarray}
 &  & \frac{1}{\beta^{2}A^{2}}\sum_{p_{1},p_{2}}\left[4P_{p_{1}-p_{2}p_{1}-p_{2}}u_{\mathbf{k}_{1}}^{2}\upsilon_{\mathbf{k}_{2}}^{2}
 +4P_{p_{1}-p_{1}p_{2}-p_{2}}u_{\mathbf{k}_{1}}\upsilon_{\mathbf{k}_{1}}u_{\mathbf{k}_{2}}\upsilon_{\mathbf{k}_{2}}
 +2P_{p_{1}p_{2}p_{1}p_{2}}\left(u_{\mathbf{k}_{1}}^{2}u_{\mathbf{k}_{2}}^{2}+\upsilon_{\mathbf{k}_{1}}^{2}\upsilon_{\mathbf{k}_{2}}^{2}\right)\right]
 G_{p_{1}}G_{p_{2}}\\
 & = & \frac{1}{4\pi^{2}\beta_{A}}\left(\left\langle |\gamma_{\mathbf{k}}|u_{\mathbf{k}}\upsilon_{\mathbf{k}}\right\rangle _{\mathbf{k}}\right)^{2}
 +\frac{1}{8\pi^{2}}\left\langle \beta_{\mathbf{k}_{1}-\mathbf{k}_{2}}\left(u_{\mathbf{k}_{1}}^{2}+\upsilon_{\mathbf{k}_{1}}^{2}\right)
 \left(u_{\mathbf{k}_{2}}^{2}+\upsilon_{\mathbf{k}_{2}}^{2}\right)\right\rangle _{\mathbf{k}_{1},\mathbf{k}_{2}};\label{eq:F2a}\end{eqnarray}
 the contribution from the diagram {}``$\ominus$'' equals

\begin{eqnarray}
 &  & -\frac{12a_{\mu}}{\beta_{A}\beta^{2}A^{2}}\sum_{p_{1},p_{2},p_{3}}\left|{\textstyle \prod_{p_{1}p_{2}p_{3}}^{\prime}}-{\textstyle \prod_{p_{1}p_{2}p_{3}}}\right|^{2}G_{p_{1}}G_{p_{2}}G_{p_{3}}\nonumber \\
 & = & -\frac{3}{\pi^{2}}<\left|{\textstyle \prod_{\mathbf{k}_{1}\mathbf{k}_{2}\left\langle -\mathbf{k}_{1}-\mathbf{k}_{2}\right\rangle }^{\prime}}-{\textstyle \prod_{\mathbf{k}_{1}\mathbf{k}_{2}\left\langle -\mathbf{k}_{1}-\mathbf{k}_{2}\right\rangle }}\right|^{2}\frac{1}{E(\mathbf{k}_{1})+E(\mathbf{k}_{2})+E(\left\langle \mathbf{k}_{1}+\mathbf{k}_{2}\right\rangle )}>_{\mathbf{k}_{1},\mathbf{k}_{2}};\label{eq:F2b}\end{eqnarray}
 and the contribution from the diagram {}``$\bigcirc\!$-$\!\bigcirc$
''equals

\begin{eqnarray}
 &  & -\frac{8a_{\mu}}{\beta_{A}\beta^{2}A^{2}}\sum_{p_{1},p_{2},p_{3}}\left(\Lambda_{p_{1}p_{2}p_{1}}-\Lambda_{p_{1}p_{2}p_{1}}^{\prime}\right)\left(\Lambda_{p_{3}p_{2}p_{3}}^{*}-\Lambda_{p_{3}p_{2}p_{3}}^{\prime*}\right)G_{p_{1}}G_{p_{2}}G_{p_{3}}\nonumber \\
 & = & -\frac{1}{4\pi^{2}\beta_{A}}\left(\left\langle \left[\beta_{\mathbf{k}}(u_{\mathbf{k}}^{2}+\upsilon_{\mathbf{k}}^{2})-|\gamma_{\mathbf{k}}|u_{\mathbf{k}}\upsilon_{\mathbf{k}}\right]\right\rangle _{\mathbf{k}}\right)^{2}.\label{eq:F2c}\end{eqnarray}
 We have set the temperature $T$ to zero and the area $A$ to infinity
in the end. The notation $\left\langle \cdots\right\rangle
_{\mathbf{k}_{1},\mathbf{k}_{2}}\equiv\int_{BZ}\frac{d^{2}\mathbf{k}_{1}}{2\pi}\int_{BZ}\frac{d^{2}\mathbf{k}_{2}}{2\pi}$,
means average over the Brillouin zone, and$\left\langle
\mathbf{k}_{1}+\mathbf{k}_{2}\right\rangle $ represents the reduced
wave vector in the Brillouin zone. Obviously,
$\frac{\partial}{\partial a_{\mu}}\mathcal{F}_{2}(a_{\mu})=0$, and
hence the two-loop correction to the atom number density is zero,
$n_{2}(a_{\mu})=0$. By Taylor expansion, one can confirm that each
of the three contributions in Eqs. \eqref{eq:F2a}, \eqref{eq:F2b}
and \eqref{eq:F2c} has infrared divergences. However, all the
divergences are exactly canceled if they are summed up. By numerical
integration, we find that

\begin{equation}
\mathcal{F}_{2}(a_{\mu})=0\end{equation} It is amazing that it
precisely amounts to zero, and the reason is still under
investigations.

\subsection{The filling fraction}

Up to two-loop, the free energy density has been obtained as
\begin{equation}
\mathcal{F}_{0+1+2}(a_{\mu})=-\frac{a_{\mu}^{2}}{4\beta_{A}}+\frac{a_{\mu}}{4\pi\beta_{A}}\left\langle
E(\mathbf{k})\right\rangle
_{\mathbf{k}},\label{eq:f_pert}\end{equation} and the atom number
density equals\begin{equation}
n_{0+1+2}(a_{\mu})=\frac{a_{\mu}}{2\beta_{A}}-\frac{1}{4\pi\beta_{A}}\left\langle
E(\mathbf{k})\right\rangle _{\mathbf{k}}.\end{equation} The filling
fraction, $n_{0+1+2}/n_{\upsilon}$, is given by\begin{equation}
\nu=\frac{\pi}{\beta_{A}}a_{\mu}-\frac{1}{2\beta_{A}}\left\langle
E(\mathbf{k})\right\rangle
_{\mathbf{k}}.\label{eq:nu_pert}\end{equation} Using the filling
fraction $\nu$ as a parameter, we have the free energy
density\begin{equation}
\mathcal{F}_{0+1+2}(\nu)=-\frac{\beta_{A}}{4\pi^{2}}\nu^{2}+\frac{1}{16\pi^{2}\beta_{A}}\left(\left\langle
E(\mathbf{k})\right\rangle _{\mathbf{k}}\right)^{2},\end{equation}
and numerically it is
$\mathcal{F}_{0+1+2}(\nu)=-0.029\nu^{2}+0.00062$. To quantify the
significance of quantum fluctuations, we calculate the ratio
$|\frac{\mathcal{F}_{1+2}}{\mathcal{F}_{0}}|$ , given by
$\frac{0.107}{\nu+0.0536}$, from which one can infer that at very
large filling fractions, the quantum fluctuation effect is
negligible and the mean field theory is adequate; at very small
filling fractions, the quantum correction is even greater than the
mean field contribution and the mean field vortex-lattice will be
unstable due to drastic fluctuations. The quantum melting of the
vortex-lattice has been intensively studied
\cite{key-25,key-26,key-27}. In the next section, we will locate the
spinodal point of the vortex-lattice state, i.e. the terminal point
of the meta-stable vortex lattice.

\section{Gaussian variational calculation\label{sec:Gaussian variational calculation}}

In this section we are going to study the model by the Gaussian
variational method. As in section \ref{sec:long-range correlation},
we set

\begin{equation}
\Psi(\mathbf{r},\tau)=\sqrt{n_{c}}\varphi(\mathbf{r})+\psi(\mathbf{r},\tau),\end{equation}
where $n_{c}$ is a real number given by minimizing the free energy
and $\psi(\mathbf{r},\tau)$ is expanded as

\begin{equation}
\psi(\mathbf{r},\tau)=\frac{1}{\sqrt{A\beta}}\sum_{\mathbf{k}\in
BZ}\sum_{m}\frac{O_{\mathbf{k}m}+iA_{\mathbf{k}m}}{\sqrt{2}}\varphi_{\mathbf{k}}(\mathbf{r})e^{-\frac{i}{2}\theta_{\mathbf{k}}}e^{-i\omega_{m}\tau},\end{equation}
where $O_{\mathbf{k}m}^{\ast}=O_{\mathbf{-k}-m}$ ,
$A_{\mathbf{k}m}^{\ast}=A_{-\mathbf{k}-m}$. In powers of
$O_{\mathbf{k}m}$ and $A_{\mathbf{k}m}$, we divide the action
$S[\Psi^{\ast}\Psi]$ into four parts
$S_{c}$,$S_{2}$,$S_{3}$,$S_{4}$.
\begin{eqnarray}
S_{c} & = & A\beta\left(-a_{\mu}n_{c}+\beta_{A}n_{c}^{2}\right),\nonumber \\
S_{2} & = & \frac{1}{2}\sum_{p}\left(\begin{array}{cc} O_{-p} &
A_{-p}\end{array}\right)\left(\begin{array}{cc}
E_{\mathbf{k}}^{O} & \omega_{m}\\
-\omega_{m} &
E_{\mathbf{k}}^{A}\end{array}\right)\left(\begin{array}{c}
O_{p}\\
A_{p}\end{array}\right).\end{eqnarray} where $O_{p}\equiv
O_{\mathbf{k}m}$ , $A_{p}\equiv A_{\mathbf{k}m}$ , and
\begin{eqnarray}
E_{\mathbf{k}}^{O} & = & -a_{\mu}+4n_{c}\beta_{\mathbf{k}}+2n_{c}|\gamma_{\mathbf{k}}|,\nonumber \\
E_{\mathbf{k}}^{A} & = &
-a_{\mu}+4n_{c}\beta_{\mathbf{k}}-2n_{c}|\gamma_{\mathbf{k}}|.\end{eqnarray}
Following the standard Gaussian variational procedure
\cite{key-38,key-39,key-40}, we first define the Gaussian
variational kernel \begin{equation}
G[\varepsilon^{O},\varepsilon^{A}]=\frac{1}{2}\sum_{p}\left(\begin{array}{cc}
O_{-p} & A_{-p}\end{array}\right)\left(\begin{array}{cc}
\varepsilon_{\mathbf{k}}^{O} & \omega_{m}\\
-\omega_{m} &
\varepsilon_{\mathbf{k}}^{A}\end{array}\right)\left(\begin{array}{c}
O_{p}\\
A_{p}\end{array}\right),\end{equation} where
$\varepsilon_{\mathbf{k}}^{O}$ and $\varepsilon_{\mathbf{k}}^{A}$
are real variational parameters. The grand-canonical partition
function can be written as \begin{eqnarray}
\mathcal{Z} & = & \int\mathcal{D}[O\, A]e^{-G}e^{G-\left(S_{c}+S_{2}+S_{3}+S_{4}\right)}\nonumber \\
 & = & e^{-S_{c}}\int\mathcal{D}[O\, A]e^{-G}\sum_{n=0}^{\infty}\frac{(-1)^{n}}{n!}\left(S_{2}-G+S_{3}+S_{4}\right)^{n},\end{eqnarray}
and the free energy density, $\mathcal{F}[n_{c},G]$, is given
by\begin{equation} \frac{1}{\beta
A}\left[S_{c}-\ln\int\mathcal{D}[O\,
A]e^{-G}-\sum_{n=1}^{\infty}\frac{(-1)^{n}}{n!}\left\langle
\left(S_{2}-G+S_{3}+S_{4}\right)^{n}\right\rangle
_{G}\right],\end{equation} where $\left\langle \right\rangle _{G}$
denotes the sum of all the connected Feynman diagrams with $G$ as a
propagator. Truncated to the first order, the free energy density
$\mathcal{F}[n_{c},G]$ takes the form\begin{equation} \frac{1}{\beta
A}\left[S_{c}-\ln\int\mathcal{D}[O\, A]e^{-G}+\left\langle
S_{2}-G+S_{3}+S_{4}\right\rangle _{G}\right].\end{equation}

\begin{figure}
\includegraphics[scale=0.8]{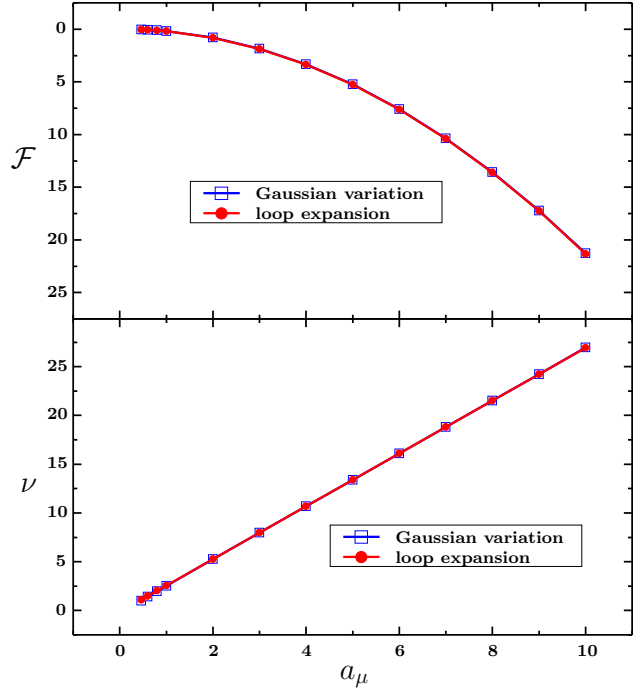}
\caption{The free energy density $\mathcal{F}$ and the filling
fraction $\nu$ are shown. The blue square symbol represents the data
obtained by Gaussian variational method, and the red circle symbol
represents the result of the perturbation theory up to
one-loop.}
\label{fig:free and fill}
\end{figure}

\begin{figure}
\includegraphics[scale=0.8] {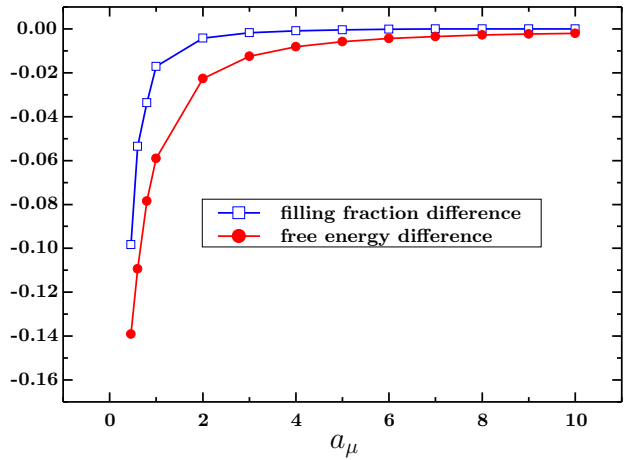}
\caption{The relative difference, that is the ratio of the result of
the Gaussian variational method minus that of the perturbation
theory to the result of the perturbation theory, is shown. The blue
square symbol represents the relative difference of the filling
fraction, and the red circle symbol represents that of the free
energy density.}
\label{fig:relative-difference}
\end{figure}

Minimizing $\mathcal{F}[n_{c},G]$ with respect to
$\varepsilon_{\mathbf{k}}^{O}$,$\varepsilon_{\mathbf{k}}^{A}$ and
$n_{c}$ leads to the following coupled equations \begin{eqnarray}
\varepsilon_{\mathbf{k}}^{O} & = & E_{\mathbf{k}}^{O}+\frac{1}{2\pi}\left\langle \beta_{\mathbf{k}-
\mathbf{k^{\prime}}}\left(\sqrt{\frac{\varepsilon_{\mathbf{k}^{\prime}}^{A}}{\varepsilon_{\mathbf{k}^{\prime}}^{O}}}
+\sqrt{\frac{\varepsilon_{\mathbf{k}^{\prime}}^{O}}{\varepsilon_{\mathbf{k}^{\prime}}^{A}}}\right)\right\rangle _{\mathbf{k}^{\prime}}
+\frac{|\gamma_{\mathbf{k}}|}{4\pi\beta_{A}}\left\langle |\gamma_{\mathbf{k}^{\prime}}|\left(\sqrt{\frac{\varepsilon_{\mathbf{k}^{\prime}}^{A}}
{\varepsilon_{\mathbf{k}^{\prime}}^{O}}}-\sqrt{\frac{\varepsilon_{\mathbf{k}^{\prime}}^{O}}
{\varepsilon_{\mathbf{k}^{\prime}}^{A}}}\right)\right\rangle _{\mathbf{k}^{\prime}},\nonumber \\
\varepsilon_{\mathbf{k}}^{A} & = & E_{\mathbf{k}}^{A}+\frac{1}{2\pi}\left\langle \beta_{\mathbf{k}
-\mathbf{k^{\prime}}}\left(\sqrt{\frac{\varepsilon_{\mathbf{k}^{\prime}}^{A}}{\varepsilon_{\mathbf{k}^{\prime}}^{O}}}
+\sqrt{\frac{\varepsilon_{\mathbf{k}^{\prime}}^{O}}{\varepsilon_{\mathbf{k}^{\prime}}^{A}}}\right)\right\rangle _{\mathbf{k}^{\prime}}
-\frac{|\gamma_{\mathbf{k}}|}{4\pi\beta_{A}}\left\langle |\gamma_{\mathbf{k}^{\prime}}|\left(\sqrt{\frac{\varepsilon_{\mathbf{k}^{\prime}}^{A}}
{\varepsilon_{\mathbf{k}^{\prime}}^{O}}}-\sqrt{\frac{\varepsilon_{\mathbf{k}^{\prime}}^{O}}{\varepsilon_{\mathbf{k}^{\prime}}^{A}}}\right)
\right\rangle _{\mathbf{k}^{\prime}},\\
n_{c} & = &
\frac{a_{\mu}}{2\beta_{A}}-\frac{1}{8\pi\beta_{A}}\left\langle
\sqrt{\frac{\varepsilon_{\mathbf{k}}^{O}}{\varepsilon_{\mathbf{k}}^{A}}}\left(2\beta_{\mathbf{k}}-|\gamma_{\mathbf{k}}|\right)\right\rangle
_{\mathbf{k}}-\frac{1}{8\pi\beta_{A}}\left\langle
\sqrt{\frac{\varepsilon_{\mathbf{k}}^{A}}{\varepsilon_{\mathbf{k}}^{O}}}\left(2\beta_{\mathbf{k}}+|\gamma_{\mathbf{k}}|\right)\right\rangle
_{\mathbf{k}},\nonumber \end{eqnarray} which can be solved
numerically. Note that the effective chemical potential $a_{\mu}$ is
the only physical parameter. Beginning with $a_{\mu}=10$ and slowly
lowering it, we find that these equations cease admitting a solution
when $a_{\mu}<0.47$, which indicates that the system can no longer
support a vortex-lattice solution and hence it is the spinodal
point. The atom number density can be obtained as the partial
derivative of the free energy density with respect to the effective
chemical potential and then the filling fraction is easily
calculated. At the spinodal point, the filling fraction equals
$1.1$, smaller than that at the quantum melting point
\cite{key-25,key-26,key-27}. In Fig. \ref{fig:free and fill} the
free energy density and the filling fraction obtained by the
Gaussian variational method are shown, together with those obtained
by the perturbation theory and expressed in Eqs. \eqref{eq:f_pert}
and \eqref{eq:nu_pert}. To take a closer look at the very small
differences of the results obtained by the two approaches, the
relative difference, that is the ratio of the result of the Gaussian
variational method minus that of the perturbation theory to the
result of the perturbation theory, is plotted in Fig.
\ref{fig:relative-difference}. From the two figures, we see that the
differences are very small and the higher the filling fraction, the
smaller the relative difference. It reveals the fact that in the
vortex-lattice state the perturbation theory and the Gaussian
variational method coincide very well, and as the filling fraction
gets higher, the two methods get closer and finally both are
identical to the mean field theory in the large filling fraction
limit.

\section{Summary\label{sec:Summary}}

We calculate the condensate density perturbatively to show that no
condensate is present in the thermodynamic limit. By calculating the
single-particle correlation function and the density fluctuation
correlation function, we obtain the algebraic decay exponent. We
calculate the free energy density to two-loop and show the
cancelation of the two-loop infrared divergences. The atom number
density distribution to one-loop is obtained, which illustrates the
quantum fluctuation effects to smooth away the mean field
vortex-lattice. By the non-perturbative Gaussian variational method,
we locate the spinodal point of the vortex-lattice, where the
filling fraction, $\nu_{s}$, is numerically obtained to be about
$1.1$, lower than the quantum melting point obtained by various
approaches \cite{key-25,key-26,key-27}. Between the spinodal point
and the melting point is the meta-stable vortex-lattice state. From
Fig. \ref{fig:local density} we find that at the one-loop level,
near the melting point where the filling fraction equals $6\sim10$,
the atom number density still retains the lattice configuration,
while near the spinodal point where the filling fraction is about
$1.1$, the lattice is almost smoothed away.

In order to determine the melting point accurately, we shall obtain
the free energy density of the vortex liquid phase in the future.
The study of the vortex liquid will be our focus in the future
studies.

\section{Acknowledgement}

We shall thank Prof. Baruch Rosenstein and Prof. Zhongshui Ma for
stimulating discussions. The work is supported by National Science
Foundation (\#10974001) and the Fundamental Research Funds for the
Central Universities.

\end{document}